\documentclass[sigconf,screen]{acmart}
\usepackage{verbatim}
\usepackage{booktabs}
\usepackage{hyperref}
\hypersetup{colorlinks,allcolors=black} 

\hypersetup{
    colorlinks=true,
    linkcolor=blue,
    filecolor=magenta,      
    urlcolor=cyan,
    pdftitle={Overleaf Example},
    pdfpagemode=FullScreen,
    }
\AtBeginDocument{%
  }

\copyrightyear{2025} 
\acmYear{2025} 
\setcopyright{cc}
\setcctype{by}
\acmConference[FSE Companion '25]{33rd ACM International Conference on the Foundations of Software Engineering}{June 23--28, 2025}{Trondheim, Norway}
\acmBooktitle{33rd ACM International Conference on the Foundations of Software Engineering (FSE Companion '25), June 23--28, 2025, Trondheim, Norway}\acmDOI{10.1145/3696630.3728695}
\acmISBN{979-8-4007-1276-0/2025/06}

\begin{document}

\title{The Tech DEI Backlash - The Changing Landscape of Diversity, Equity, and Inclusion in Software Engineering}


\author{Sonja M. Hyrynsalmi}
\email{sonja.hyrynsalmi@lut.fi}
\orcid{0000-0002-1715-6250}
\affiliation{%
  \institution{LUT University}
  \city{Lahti}
  \country{Finland}
}

\author{Mary Sánchez-Gordón}
\email{mary.sanchez-gordon@hiof.no}
\orcid{0000-0002-5102-1122}
\affiliation{%
  \institution{Østfold University College}
  \city{Halden}
  \country{Norway}}

\author{Anna Szlavi}
\email{anna.szlavi@ntnu.no}
\orcid{0000-0002-4041-6217}
\affiliation{%
  \institution{Norwegian University of Science and Technology}
  \city{Trondheim}
  \country{Norway}
}

\author{Letizia Jaccheri}
\email{letizia.jaccheri@ntnu.no}
\orcid{0000-0002-5547-2270}
\affiliation{%
 \institution{Norwegian University of Science and Technology}
 \city{Trondheim}
 \country{Norway}}



\begin{abstract}
Not long ago, Diversity, Equity, and Inclusion (DEI) initiatives were a top priority for leading software companies. However, in a short period, a wave of backlash has led many firms to re-assess their DEI strategies. Responding to this DEI backlash is crucial in academic research, especially because, currently, little scholarly research has been done on it. In this paper, therefore, we have set forth the following research question (RQ): 
"How have leading software companies changed their DEI strategies in recent years?" 
Given the novelty of the RQ and, consequently, the lack of scholarly research on it, we are conducting a grey literature study, examining the current state of DEI initiatives in 10 leading software companies. Based on our analysis, we have classified companies into categories based on their shift in commitment to DEI. We can identify that companies are indeed responding to the backlash by rethinking their strategy, either by reducing, increasing, or renaming their DEI initiatives. In contrast, some companies keep on with their DEI strategy, at least so far, despite the challenging political climate.  To illustrate these changes, we introduce the DEI Universe Map, a visual representation of software industry trends in DEI commitment and actions.
\end{abstract}




\maketitle

\section{Introduction}
In recent years, Diversity, Equity, and Inclusion (DEI) actions and initiatives have taken various forms in the corporate world, from inclusive leadership training through employee resource groups to pay equity audits, as stakeholders have demanded transparency and commitment to DEI practices ~\cite{thompson2024coping}. Software companies that develop, deliver, or distribute software systems have been key players in DEI initiatives. For example, the research by Moises and Jaccheri~\cite{MoisesJaccheri2024} and Szlavi and Guedes~\cite{10.1007/978-3-031-35699-5_25} revealed how big technology companies (Google, Amazon, Meta, Apple, Microsoft) report thoughtfully inclusion and diversity actions in their software product and service design.

However, a recent study based on data from 1,000 companies with DEI programs found that 1 in 8 organizations will reduce their DEI commitments in 2025, citing the political climate as the primary reason, with more joining every day~\cite{Resume2025}. The political climate surrounding DEI initiatives has shifted dramatically in the US, particularly since Donald Trump began his second presidential term, with increased scrutiny on corporate DEI efforts, legislative actions limiting DEI funding, and growing pressure from stakeholders advocating for its reduction. Major software firms such as Amazon, Meta (Facebook), Alphabet (Google), and Accenture, once at the forefront of DEI advocacy, have been reducing their DEI teams or reframing DEI initiatives under some broader topics.~\cite{MurrayBohannon2025, AllenFischer2025, AllenFischer2025b, Zuckerberg2025}. This situation has been broadly described as a "DEI backlash"~\cite{thompson2024coping}.

The terms "backlash" or "blowback" refers to a strong negative reaction to a movement or policy, this time around DEI~\cite{hill2009incorporating}. Still, DEI is not new; in fact, the management of DEI actions in companies and their impact on companies have been researched for decades~\cite{geber1990managing, kossek1996managing}. Furthermore, there have already been backlash periods before, as well as reported challenges when implementing DEI initiatives at the company level~\cite{kidder2004backlash, bezrukova2012reviewing}. However, what differentiates this backlash from the previous instance of DEI challenges is the speed of the change~\cite{thompson2024coping, sitzmann2024don}. Moreover, the current backlash of DEI is labeled with terms such as "woke" and "wokeness" as a way to justify it, even though the true meaning of "woke" remains vague~\cite{prasad2024critiquing}.

As software companies develop technologies that shape societies around the world, monitoring, analyzing, and responding to this DEI backlash is crucial in academic research as well. Understanding the current state of DEI in leading software companies helps us to provide a clearer picture of the evolving landscape and identify whether these shifts represent temporary adjustments or a fundamental transformation in corporate values. For that reason, we approach this topic via one research question: 

\begin{itemize}
    \item RQ: How have leading software companies changed their DEI strategies in recent years?
\end{itemize}

To answer this research question, we conducted a grey literature review of recent news and reports from the 10 leading software companies on the Forbes 2024 Global 2000 list. 

\section{Background}

\begin{figure*}[tbh]

\includegraphics[width=0.43\textwidth]{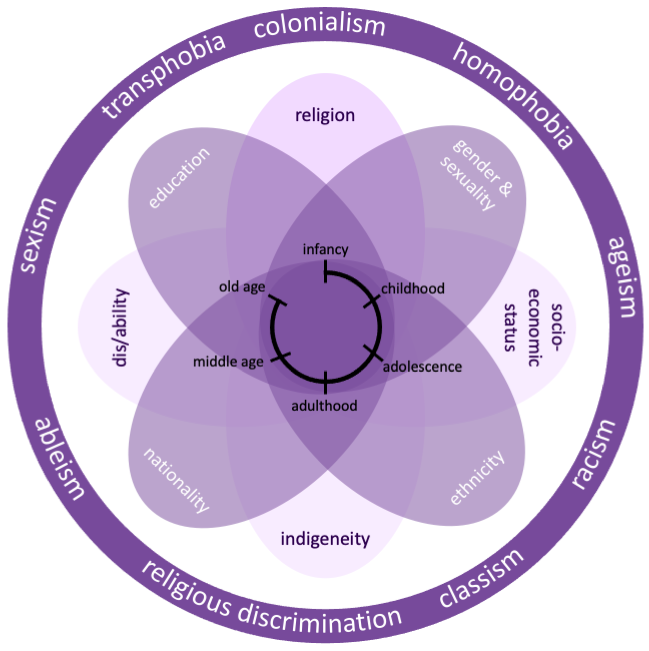}

  \caption{Intersectionality wheel, provided by UN Women~\cite{un-intersectionality} displays several dimensions of differences and inequalities.}
  \label{intersectionality-fig}
\end{figure*}

During the past two decades, big technology companies have expanded their role beyond innovation and market dominance to embrace broader social responsibility. As stakeholders demand transparency and ethical practices, these companies have invested in DEI. Early advocates like Ben \& Jerry's highlight the relevance of social justice to corporate reputation~\cite{Lim2021EffectsOI, CiszekLogan2018}. DEI has gained ground, especially in the tech industry, prompting companies to adapt their discourse and practices and software engineering researchers to think about how to revise theories and frameworks to incorporate these ideas~\cite{DamianD2024DEIinSE}. Previous reports from major tech companies such as Alphabet (Google), Amazon, Meta (Facebook), Apple, and Microsoft have been demonstrating their commitment to building a better society~\cite{wichowski2020information,10.1007/978-3-031-35699-5_25}.

Feminism, both as a theoretical perspective and a social movement, aims to eliminate sexist inequality and oppression~\cite{bell2020}. Intersectionality~\cite{davis2008intersectionality,choo2010practicing} highlights the systemic power dynamics that arise from the interaction of various social differences across individual, institutional, cultural, and societal spheres. Figure \ref{intersectionality-fig} shows the Intersectionality wheel by UN Women~\cite{un-intersectionality}, with the overlapping identity segments and the type of discrimination that can arise from those segments. Intersectionality studies~\cite{rodriguez16, bell2020, choo2010practicing} have not addressed software engineering specifically in their approach, and the direct intersectional approach has been underrepresented in SE research and only recently gaining more attention~\cite{szlavi2024integrating, sanchez2021framework}. 

Previous research has examined how leading technology companies incorporate diversity, equity, and inclusion (DEI) into their design principles and corporate reporting. Szlavi and Guedes~\cite{10.1007/978-3-031-35699-5_25} investigated company websites to determine whether inclusivity and DEI were explicitly mentioned as guiding principles and, if so, how these commitments were framed. Similarly, Moises de Souza and Jaccheri~\cite{MoisesJaccheri2024} analyzed in their grey literature review major technology companies’ corporate reports—specifically Corporate Social Responsibility (CSR), Diversity, Equity, and Inclusion (DEI), and Environmental, Social, and Governance (ESG) reports—to assess whether and how they reported diversity and inclusion efforts in their product and service design. 



\textit{The shift in the attitudes towards DEI} 

Understanding the opposition of advantaged groups to diversity, equity, and inclusion policies has always been at the core of DEI research. For example, Iyer~\cite{iyer2022understanding} has investigated how DEI policies often generate opposition from advantaged groups due to perceived threats. These threats (resource threat, symbolic threat, and identity threat) can shape resistance to DEI efforts in different ways. Resource threat arises when advantaged groups feel that increasing diversity may limit their opportunities, such as through preferential hiring or promotion policies. A symbolic threat emerges when DEI initiatives are perceived as challenging existing norms, values, or meritocratic principles. Identity threat, in turn, stems from the fear that one's group may lose its centrality or status in key societal domains, such as STEM fields or leadership positions.

Right now, DEI initiatives are facing strong opposition, especially in the US, which may also influence the global mindset. On January 21, 2025, President Trump signed an Executive Order titled “Ending Illegal Discrimination and Restoring Merit-Based Opportunity” (the “January 21 DEI Order”). Furthermore, on 20 January 2025, President Trump signed two DEI-related executive orders titled “Defending Women From Gender Ideology Extremism and Restoring Biological Truth to the Federal Government" (the "Gender Order") and "Ending Radical and Wasteful Government DEI Programs and Preferencing: (the "DEI Order 20 January")~\cite{High2025, WhiteHouse2025a, WhiteHouse2025b}.  

Furthermore, the backlash in DEI also affects associations that work on DEI initiatives and not only companies. Recently, organizations such as Girls in Tech and Women Who Code have been forced to shut down or rebrand, in order to survive amid growing resistance to diversity efforts, as funding has declined significantly~\cite{WashingtonPost2024}.

In addition, research around DEI topics is facing challenges, especially in the US. For example, the "Gender Order"~\cite{WhiteHouse2025b} states that \textit{"each agency and all Federal employees shall enforce laws governing sex-based rights, protections, opportunities, and accommodations to protect men and women as biologically distinct sexes."} Furthermore, it mandates that \textit{"agencies shall remove all statements, policies, regulations, forms, communications, or other internal and external messages that promote or otherwise inculcate gender ideology, and shall cease issuing such statements, policies, regulations, forms, communications, or other messages."} Additionally, the order instructs agencies to ensure that \textit{"federal funds shall not be used to promote gender ideology"} and that all grant conditions and funding mechanisms are aligned with this directive. Although the order does not explicitly mention diversity or DEI initiatives, it directs agencies to review and rescind any materials that include gender-identity-based policies. This may indirectly impact broader DEI programs, particularly those focused on gender inclusion and intersectional approaches that address overlapping forms of marginalization, such as race, ethnicity, and disability. However, the order does not appear to directly target DEI initiatives beyond those related to gender identity.

\subsection{The Shift to Standards} 
DEI has become increasingly relevant in corporate agendas, and major tech corporations are no exception~\cite{MoisesJaccheri2024}. As it has evolved into stakeholder requirements, companies are required to adopt more structured approaches and demonstrate measurable progress. As a result, multiple national standards on DEI have emerged. For example, the British Standard BS 76005:2017 “Valuing people through diversity and inclusion. Code of practice for organizations”, the Norwegian Standard NS 1120:2018 “Diversity Management System” or the Italian Standard UNI/PdR 125:2022 “Gender Equality Management System”. In this trend, the ISO 30415 - International Standard about “Human resource management — Diversity and inclusion” standard, was published in 2021.

The ISO 30415 standard has established the first global framework for implementing inclusion as a structured change management process~\cite{ISO30415:2021}. It provides guidance in two key areas: Defining stakeholder needs and organizing a diverse portfolio of inclusion-focused services. The standard outlines the service types that can be delivered to support DEI initiatives: i) Training: A wide range of education methods tailored to meet the needs of identifiable stakeholders, iii) Data Extraction: A variety of data collection activities, both voluntary and involuntary, and iii/iv) Internal/External Infrastructure: Any and all internal/external resources established to support and accommodate stakeholders. In addition, four key operational areas for implementing  DEI initiatives within an organization are identified: i) Governance: DEI services provided to management, ii) Product Delivery: DEI services integrated into in/tangible goods, iii) Human Resources: DEI services directed at employees and iv) Supplier Diversity: DEI services extended to the supply chain and other organizational stakeholders. Governance and product delivery processes should be implemented across all types of organizations. 

The standard is relevant to the following United Nations Sustainable Development Goals (SDGs): 5) Gender Equality, 8) Decent Work and Economic Growth, 9) Industry, Innovation and Infrastructure, 10) Reduced Inequality, and is guided by the principles of human rights at work.

\section{Research Process}
\begin{table*}[h]
    \centering
    \begin{tabular}{rllc}
        \toprule
        \textbf{Rank} & \textbf{Company} & \textbf{Country} & \textbf{Market Value*} \\
        \midrule
        6  & Amazon (AWS)        & United States & 1,922.1 \\
        8  & Microsoft           & United States & 3,123.1 \\
        10 & Alphabet (Google)   & United States & 2,177.7 \\
        12 & Apple               & United States & 2,911.5 \\
        24 & Meta Platforms      & United States & 1,197.0 \\
        41 & Alibaba Group       & China         & 200.8   \\
        77 & Oracle              & United States & 339.4   \\
        85 & IBM                 & United States & 155.3   \\
        158 & Salesforce         & United States & 234.6   \\
        230 & Adobe              & United States & 208.5   \\
        \bottomrule
    \end{tabular}
    \caption{Top 10 Software Companies from Forbes 2024 Global 2000 list used in this research. Note: *(USD B)}
    \label{tab:top_software_companies}
\end{table*}

To answer our research questions, we conducted a Grey Literature Review (GLR) ~\cite{garousi2019guidelines, adams2017shades} to investigate how leading software companies have recently adjusted their DEI strategies. We chose GLR because traditional peer-reviewed academic literature does not cover the recent shifts in companies' DEI strategies. Also, to undercover the information about changes in companies' DEI strategies, it is useful to investigate grey literature such as memos, emails, web page changes, or news from the topic. 

We focus on 10 leading software companies on the Forbes 2024 Global 2000 list\footnote{https://www.forbes.com/lists/global2000/} (Table \ref{tab:top_software_companies}). To define a software company, we mean companies that primarily focus on developing, distributing, or providing software products as a core business model. We chose this focus because these companies are the key players in delivering diverse, equal, and inclusive software systems worldwide. At this time, we excluded hardware companies, such as chip component companies, to focus more on software products and services. 

\subsection{Data collection and analysis}

We conducted two grey literature data collection sets: Data 1 concerns datasets about DEI key numbers in companies, and Data 2 focuses on news articles and other grey literature about recent DEI actions.  

Given the limited literature on this topic, for Data 1 we explored raw databases to provide context related to employee numbers (\href{http://www.stockanalysis.com}{StockAnalysis}\footnote{https://stockanalysis.com/}), layoffs (\href{http://www.layoffstracker.com}{LayoffsTracker}\footnote{https://layoffstracker.com/}), and Diversity and Inclusion (D{\&}I) ratings  (\href{http://www.glassdoor.com}{Glassdoor}\footnote{https://www.glassdoor.com/}). The first database provides the annual number of employees according to filings submitted to the U.S. Securities and Exchange Commission (SEC). Data is also manually gathered from company press releases, IPO filings, and other official sources. LayoffsTracker is a free service that tracks layoffs across companies in the US and worldwide, while Glassdoor is a website where current and former employees anonymously review companies. 

To systematically collect relevant grey literature regarding DEI strategy changes in Data 2, we conducted separate search queries for each company using Google as our search engine. To investigate the most recent reactions potentially caused by President Trump re-election, we applied a custom date range filter in Google, filtering results to content published from January 1, onward. The data collection was made in February 2025 and we used the search string:

\begin{itemize}
    \item ("Diversity" OR "Diversity, Equity, and Inclusion" OR "DEI") AND ("[Company Name]")
\end{itemize}

To ensure reliability and relevance for news articles collected for Data 2, we followed the grey literature selection criteria of Garousi et al. ~\cite{garousi2019guidelines}. Firstly, we focused only on trusted news platforms, such as Bloomberg, Forbes, Business Insider, Axios, and CNBC. Secondly, we included company-published statements to ensure first-hand accounts of DEI strategy changes if they were available. Finally, for news articles referencing internal emails or corporate memos, we only selected those labeled as "source news" where the publication either included the full memo or email text or explicitly stated that the document had been reviewed and verified by the news organization, even if the original source's name was anonymized. In those cases where there were not any public documents, we investigated, for example, changes in the DEI-related web page sections if there were archived versions. The full list of sources included for Data 2 can be found in Appendix 1.

To analyze the Data 2 findings, we applied thematic analysis~\cite{clarke2017thematic} to identify commitment shifts in companies’ DEI strategies. The first stage of the analysis involved determining whether a company was \textit{decreasing, renaming, defending, or retaining} its DEI initiatives based on news articles, corporate memos, and other statements found in the grey literature review. Some companies exhibited multiple shifts simultaneously, indicating a more complex shift in their DEI approach. Then we categorize them to \textit{Brightest Stars (Champions), Emerging Stars (High Flyers)} and \textit{Distant Planets (On the Radar)}. This classification enabled us to capture both the intensity and direction of DEI-related actions.

Then, we combined the findings from Data 1 and Data 2 and positioned companies within the DEI Universe, a model inspired by a constellations-in-a-universe metaphor, like astronomical objects. The placement was based on the scale and depth of DEI actions, the company’s workforce size and industry influence, and their trajectory over time—whether their DEI strategy was stabilizing, shifting, or declining. The final DEI Universe visualization integrates these elements, presenting companies as bubbles, where size represents employee count and color reflects their DEI positioning. This approach provides a holistic view of how DEI strategies evolve across the software industry.

In this paper, we primarily use the term DEI (Diversity, Equity, and Inclusion) but when referring to specific company-reported metrics, we also include D{\&}I (Diversity and Inclusion), as some organizations do not explicitly differentiate between the two. For consistency, in this research, we consider both terms to represent corporate initiatives aimed at fostering diverse, inclusive, and equitable workplaces.

\section{Results}

In this study, we examine the DEI strategies of ten leading software companies selected from the Forbes Global 2000 (2024) list: Amazon, Microsoft, Alphabet, Apple, Meta Platforms, Alibaba Group, Oracle, IBM, Salesforce, and Adobe.
\subsection{Context}
The tech industry experienced hypergrowth before and during the pandemic ~\cite{Berkan2024}, as shown in Table \ref{tab:overview_software_companies}. Initially, companies grew their workforce to meet the increased demand for digital services during lockdowns, while simultaneously accelerating their efforts in artificial intelligence development. 
\begin{table*}[h]
    \centering
    \begin{tabular}{rrrrrrrrrl} 
        \toprule
        Company &\multicolumn{6}{c} \textbf{Employees  (in thousand)  } & 
        \multicolumn{1}{c} \textbf{Layoffs  } & 
        \multicolumn{2}{c} \textbf{D{\&}I rating  }  \\
        \cmidrule(lr){2-7}\cmidrule(lr){8-8}\cmidrule(lr){9-10}
        Year &
            2019 & 2020 & 2021 & 2022 & 2023 & 2024 & 
            2022-25 &
            <2025 & 
            \includegraphics[width=0.90cm]{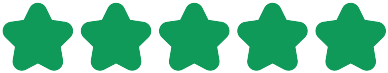}\\
        \cmidrule(lr){1-1}        
        Amazon (AWS) &
            798 & 1,298 & 1,608 & \textcolor{red}{1,541} & \textcolor{red}{1,525} & 1,551 & 37,810 & 7,927 & 3.8 \textcolor{red}{\textbf{-}}\\
        Microsoft & 
            144 & 163 & 181 & 221 & 221 & 228 & 18,042 & 12,712 & 4.3 \textcolor{red}{\textbf{-}}\\
        Alphabet (Google) &
            135 & \textcolor{red}{118} & 156 & 190 & \textcolor{red}{182} & \textcolor{red}{181} & 14,033 & 13,184 & 4.3 +\\        
        Apple & 
            137 & 147 & 154 & 164 & \textcolor{red}{161} & 164 & 200 & 10,388 & 4.2 \textcolor{red}{\textbf{-}}\\
        Meta Platforms& 
            44 & 58 & 71 & 86 & \textcolor{red}{67} & 74 & 31,701 & 6,659 & 4.0 \textcolor{red}{\textbf{-}}\\
         Alibaba Group& 
            101 & 117 & 251 & 254 & \textcolor{red}{235} & \textcolor{red}{204} & 90 & 925 & 3.4 +\\   
        Oracle & 
             136 & \textcolor{red}{135} & \textcolor{red}{132} & 143 & 164 & \textcolor{red}{159} & 3,384 & 13,608 & 4.0 \textcolor{red}{\textbf{-}}\\
        IBM & 
            352 & \textcolor{red}{345} & \textcolor{red}{282} & 288 & \textcolor{red}{282} & n/a & 5,080 & 27,968 & 4.2 \textcolor{red}{\textbf{-}}\\
        Salesforce & 
             29 & 35 & 56 & 73 & 79 & \textcolor{red}{72} & 10,590 & 8,009 & 4.2 \textcolor{red}{\textbf{-}}\\
        Adobe & 
            n/a& 22 & 25 & 29 & 29 & 30 & 100 & 1,083 & 4.2 \textcolor{red}{\textbf{-}}\\

        \bottomrule
    \end{tabular}
    \caption{Overview of  Employee Numbers, Layoffs, and D{\&}I ratings. Note: decreased (\textcolor{red}{\textbf{-}}) / increased (+) over the last year}
    \label{tab:overview_software_companies}
\end{table*}

As economic uncertainties continued beyond the lockdowns, companies re-evaluated their strategies. For instance, in 2022, Meta was among the first companies to respond to the post-pandemic economic downturn and slow revenue growth by implementing a broad hiring freeze, followed by the first round of layoffs aimed at cutting costs and increasing profitability. Table \ref{tab:overview_software_companies} shows the number of employees affected by global tech layoffs from January 2022 to February 2025. The company with the most layoffs is Amazon, followed by Meta, but Meta seems to have adopted a more aggressive approach in relation to the size of its workforce. In contrast, Apple did not follow the trend of mass layoffs and stood out as the only major tech company that prioritized its stakeholders over its shareholders, according to Çetin and Gündüz ~\cite{Berkan2024}. More recently, layoffs appear to be driven by a strategic shift toward increased automation and adoption of AI in various sectors. Layoffs can exacerbate inequalities, especially for minority groups, create barriers to re-employment, and cause job insecurity for those who remain employed after mass layoffs ~\cite{Berkan2024}. 

Finally, Table \ref{tab:overview_software_companies} shows the number of professionals who provided D{\&}I reviews and the anonymous employee ratings up to February 2025. For example, IBM’s D{\&}I rating of 4.2 out of 5 stars, based on 27,968 employee ratings. However, the average D{\&}I rating left by them decreased (\textcolor{red}{-}) over the last year. Note that the average D{\&}I rating at Alibaba and Google increased (+) over the last year.

\subsection{DEI strategies}

In our grey literature review, we discovered that companies can be differentiated in the level of their actions and in their commitment shift. We categorized companies as either \textit{defending, retaining, renaming}, or \textit{decreasing} their DEI initiatives in various ways. We further categorized companies by three key constellations: \textit{Champions, High Flyers}, and \textit{On the Radar} companies. 

\subsubsection{Brightest Stars (Champions)}
Apple, IBM, and Adobe lead the way, showing their recommitment for DEI initiatives.

\textbf{Apple:}
In our research, we identified that Apple is \textit{defending} their DEI initiatives, as in their 2025 Annual Meeting, which was held virtually on February 25, 2025, they defended several shareholder proposals that linked directly or indirectly to their DEI initiatives~\cite{Apple2025Proxy}. 

Directly towards DEI actions was a request to cease DEI efforts. In their response, Apple stated \textit{"[d]espite these obvious risks, the SFFA and Muldrow decisions and the wave of corporate DEI retreats, Apple still has an “Inclusion \& Diversity” program -- and contributing shareholder money to organizations that advance DEI.}" This is a clear sign that they decided to actively defend their DEI initiatives.

There was also a request to report on Charitable Giving, and in their response, Apple stated that "\textit{Apple has a well-established corporate donations program that follows a strict internal governance and approval process, and the proposal attempts to inappropriately restrict Apple’s ability to manage its own ordinary business operations and business strategies."}

Other requests, which the Apple board voted against in that same meeting, were 'Report on Costs and Benefits of Child Sex Abuse Material-Identifying Software \& User Privacy' and 'Report on Ethical AI Data Acquisition and Usage'~\cite{Apple2025Proxy}.

\textbf{IBM:} The Heritage Foundation~\cite{Heritage2025CorporateAccountability}, the same foundation that has used its stakeholder rights to convince Apple to drop their DEI initiatives, has also approached IBM and demanded IBM~\cite{MacLellan2025} that \textit{"[s]hareholders request the Board of Directors of IBM conduct an evaluation and issue a report within the next year, at reasonable cost and excluding confidential information, assessing how the Company’s DEI requirements for hiring/recruitment impact IBM’s risks related to discrimination against individuals based on their race, color, religion (including religious views), sex, national origin, or political views"}. 

Heritage Foundation also wrote: \textit{"IBM executives are expected to meet certain recruitment quotas of ‘underrepresented’ minorities, including various racial minority groups and women. As per Krishna’s remarks, IBM execs who fail to reach such quotas risk slashed bonuses, demotion, or termination of employment. These policies directly require IBM executives to make recruitment/promotion decisions based on race"}~\cite{IBM_Heritage_Response2024}. In this, Heritage Foundation refers to an ongoing lawsuit against IBM, claiming that IBM would discriminate against white men by using diversity quotas in recruitment~\cite{Dill2024}.

IBM has responded to the Heritage's demand that \textit{"[i]t is the policy of this organization to continue to engage in activities such as hiring, promotion and compensation of employees, without regard to race, color, religion, sex, gender, gender identity or expression, sexual orientation, national origin, caste, genetics, pregnancy, mental or physical disability, neurodivergence, age or other characteristics protected by applicable law”}. These factual misstatements are material to shareholders voting on the Proposal. A fact is material if there is a substantial likelihood that a reasonable shareholder would consider it important in deciding how to vote".~\cite{IBM_Heritage_Response2024}

\textbf{Adobe:}
In our investigation, we identified that Adobe was \textit{defending} DEI initiatives. In January, they made an additional \$5 million for its Film \& TV Fund. Adobe's Film \& TV Fund is collaborating with Group Effort Initiative (GEI) to provide training courses for aspiring filmmakers and professionals from underrepresented communities, focusing on career development in the industry~\cite{Martinet2025}. 

They also have continued collaboration with The Annenberg Inclusion Initiative, which is the leading think tank in the world studying diversity and inclusion in entertainment. In January 2025, they announced that they are revealing together Dr. Stacy L. Smith and the USC Annenberg Inclusion Initiative the Inclusion List. In this list, they rank: \textit{"the top films, distributors, directors, and editors on a point system that accounts for inclusion among the cast – both leading and all speaking characters – and crew across five indicators (gender, race/ethnicity, LGBTQ+, disability, and age)"}~\cite{Martinet2025b}.

In January 2025, Adobe was also looking to hire a Director, DEI Advisory\footnote{https://www.linkedin.com/jobs/view/director-dei-advisory-at-adobe-3903002000/}, but our investigation does not provide a conclusion if that recruitment process was finalized and if the final title of the person would indeed be so DEI-focused. 

Furthermore, they also state in their webpage: \textit{"At Adobe, we believe that when people feel respected and included, they can be more creative, innovative, and successful. While we have more work to do to advance diversity and inclusion, we’re investing to move our company and industry forward."}\footnote{https://www.adobe.com/diversity.html}

\subsubsection{Emerging Stars (High Flyers)} Microsoft, Alibaba, Oracle, and Salesforce show potential and commitment but need more impact or action to reach the top tier.

\textbf{Microsoft:}
Microsoft declared in their Global Diversity and Inclusion Report 2024, published in October 2024~\cite{Microsoft2024}, that their \textit{"focus on diversity and inclusion has been going on for decades. Our journey continues, and the work is ongoing."} In our grey literature review, we did not see any recent outcomes around the topic besides this report, and that could give a signal that Microsoft is \textit{'retaining'} their DEI initiatives.

The biggest reason why we did not see that Microsoft was \textit{defending} their DEI initiatives was that in 2024 Microsoft laid off the whole DEI team. An email from the team leader stated that \textit{"[t]rue systems-change work associated with DEI programs everywhere are no longer business critical or smart as they were in 2020"} and that the layoffs were made because of \textit{"changing business needs"}~\cite{Stewart2024}. However, that strong declaration mentioned previously in their Global Diversity and Inclusion Report 2024~\cite{Microsoft2024} gives a signal that they are committed to their DEI initiatives. 

\textbf{Alibaba:} In our investigation, we found that Alibaba is \textit{retaining} their DEI initiatives. They have continued their Alibaba Global Initiatives and state in their mission that they aim \textit{"To share the positive impact of the new business paradigm in promoting inclusive development and to inspire and empower entrepreneurs, youth and women."}\footnote{https://agi.alibaba.com/home}

Their DEI initiatives are collected under their Alibaba Group Environmental, Social and Governance (ESG) Report, which was last time updated 2024~\cite{Alibaba2024}. There they state: \textit{"To gain the trust of our corporate partners and society, firstly, we observe ethical business practices and comply with applicable regulations, overseen by an effective, transparent, and diverse corporate governance structure"}.

\textbf{Oracle:} Oracle is \textit{retaining} and \textit{renaming} their DEI initiatives. They have not had wide publicity around the area, so we investigated their web page changes and if there have been any changes lately. 

Oracle has been making changes to its web page since January 2025. Previously, they have written in their Release Notes (Release 14.7.5.0.0) \footnote{https://docs.oracle.com/en/industries/financial-services/banking-trade-finance/14.7.5.0.0/prcug/shared-global-topic-diversity-and-inclusion.html} that: \textit{"Oracle respects and values having a diverse workforce that increases thought leadership and innovation. As part of our initiative to build a more inclusive culture that positively impacts our employees, customers, and partners, we are working to remove insensitive terms from our products and documentation."} The Google Archives found out also several web pages where the head titles mentioned 'Diversity and Inclusion', 'DE{\&}I' and 'D{\&}I'. Now, for example, Diversity and Inclusion Report link \footnote{https://www.google.com/url?sa=t{\&}source=web{\&}rct=j{\&}opi=89978449{\&}url=https://www.oracle.com/fi/human-capital-management/diversity-and-inclusion/report/} guides only to the main landing page of 'Oracle Human Capital Management (HCM)'. This gives the impression that Oracle has removed the words 'Diversity and Inclusion', 'DE{\&}I' and 'D{\&}I' from their web page and replaced them with 'Human Capital Management'. The word 'belonging' has been added for many places. 

No other public action has been taken during our investigation period, so it is not clear whether they have decreased their actions at some level or just renamed them.

\textbf{Salesforce:} In our investigation, we identified that Salesforce is \textit{defending} their DEI initiatives. In a recent interview made at World Economic Forum in Davos, January 2025, the CEO of Salesforce Marc Benioff said "[m]y job as the CEO is to kind of get their [employees'] back. If somebody's going to come after our employees or discriminate against them in any way, we'll do everything we can to help them, support them."~\cite{Lotz2025}

Salesforce has also faced lawsuits claiming that they are limiting free speech. The recent lawsuit claimed: \textit{"Digital service providers (DSPs) such as Salesforce control access to critical computer and web-related services and infrastructure that facilitate the open exchange of information. As such, these companies have unprecedented power to censor speech and have been under increasing pressure to remove religious and conservative views from the marketplace"}~\cite{NationalCenter2025}. These lawsuits go back to 2015 when Salesforce also defended their DEI policies as in Indiana, one state in USA, there was a planning for a law that would discriminate against LGBTQ+ people. The CEO Marc Benioff commented again in 2022: \textit{"Look, we have to be for equality, we have to be for dignity. We have to be for the equality and dignity of every human being, and if you're not for equality and dignity, then, you know, this is not something that I can work with and we're going to have to exit your city or your state just as we have in many places."}~\cite{LaMonica2022, CNN2022}

Salesforce presents their DEI initiatives under the term \textit{"Equality"}.\footnote{https://www.salesforce.com/company/equality/}

\subsubsection{Distant Planets (On the Radar)} Amazon, Alphabet (Google), and Meta (Facebook) are still orbiting on the edges, needing stronger efforts to move toward leadership.

\textbf{Amazon:}
From the recent news from Amazon, we are identifying that Amazon is both \textit{decreasing} and \textit{renaming} their DEI initiatives. The memo sent to the employed by Candi Castleberry, Amazon’s vice president of Inclusive eXperiences and Technology, said that Amazon is “\textit{winding down outdated programs and materials}”. It also stated that \textit{“Rather than have individual groups build programs, we are focusing on programs with proven outcomes — and we also aim to foster a more truly inclusive culture”}~\cite{Palmer2025, DayGreen2025}.

However, the reason we identified that Amazon was decreasing their DEI initiatives in the big picture is that a week before the memo mentioned above, Amazon had also eliminated specific sections titled Equity for Black people and LGBTQ+ rights from their 'Our Positions' webpage~\cite{WaytVarnhamWingfield2025}. CNBC investigated this section modification via Internet Archive’s Wayback Machine and when asking about the reasoning behind this, Amazon spokesperson Kelly Nantel told them that: “\textit{We update this page from time to time to ensure that it reflects updates we’ve made to various programs and positions.”}~\cite{Palmer2025}\footnote{https://web.archive.org/web/20241126111614/https://www.aboutamazon.com/about-us/our-positions} and did not specify how these sections would be reformed.

\textbf{Alphabet:} In our review, we identified that Alphabet, better known as Google, was \textit{decreasing} the DEI initiatives. Business Insider published a memo, in which Google's Chief People Officer Fiona Cicconi stated: \textit{"Every year, we review the programs designed to help us get there and make changes. And because we are a federal contractor, our teams are also evaluating changes to our programs required to comply with recent court decisions and U.S. Executive Orders on this topic"}~\cite{AltchekLangley2025}.

Furthermore, Google users noticed recently that Google had removed, for example, Pride Month, Black History Month, Indigenous People Month, Jewish Heritage, Holocaust Remembrance Day, and Hispanic Heritage from their calendar\footnote{https://support.google.com/calendar/thread/322570334/why-is-pride-month-removed-from-the-google-calendar?hl=en}. 

\textbf{Meta Platforms:} Findings in our research indicate that Meta is \textit{decreasing} and \textit{renaming} their DEI initiatives. Meta is behind big social media platforms such as Facebook and Instagram. Recently, the founder and CEO of Meta, Mark Zuckerberg, announced in a video message on Facebook~\cite{Zuckerberg2025} that \textit{"We're going to simplify our content policies and get rid of a bunch of restrictions on topics like immigration and gender that are just out of touch with mainstream discourse. What started as a movement to be more inclusive has increasingly been used to shut down opinions and shut out people with different ideas, and it's gone too far"}.

In addition, Meta employees got a memo in Workplace, the company's internal communications tool, about recent changes~\cite{AllenFischer2025, AllenFischer2025b}. In that memo it was emphasized that Meta had already ended representation goals for women and ethnic minorities and now informed that \textit{"[i]nstead of equity and inclusion training programs, we will build programs that focus on how to apply fair and consistent practices that mitigate bias for all, no matter your background"}.

In this memo, it was also stated that \textit{["t]he Supreme Court of the United States has recently made decisions signaling a shift in how courts will approach DEI. -- The term "DEI" has also become charged, in part because it is understood by some as a practice that suggests preferential treatment of some groups over others"}.

Furthermore, in this memo, it was announced that Meta no longer has a team focused on DEI and that their previous company's chief diversity officer is taking on a new role at Meta, focused on accessibility and engagement. This is especially a sign we identified for \textit{renaming}.

\subsection{DEI Universe}

The GLR results guided us to identify commitment shifts and a set of actions over time that revealed a dynamic and interconnected ecosystem. To better understand the evolving landscape of DEI in software engineering, this study proposes a "constellations-in-a-universe" model. Stars in the universe vary in brightness, size, color, and behavior. Some undergo rapid transformations, evolving into different types within a short period, while others remain relatively stable over a long period. 
Like celestial bodies, companies are not stationary: some may rise in prominence, while others may stagnate or decline in their DEI efforts. However, the lack of energy could be the beginning of the end of a star’s life. Moreover, the concept of gravitational pull can be inferred: companies with greater influence may exert an indirect pull on others, shaping industry-wide trends, e.g., layoffs.  

\begin{figure*}[h]
    \centering
    \includegraphics[width=0.75\textwidth]{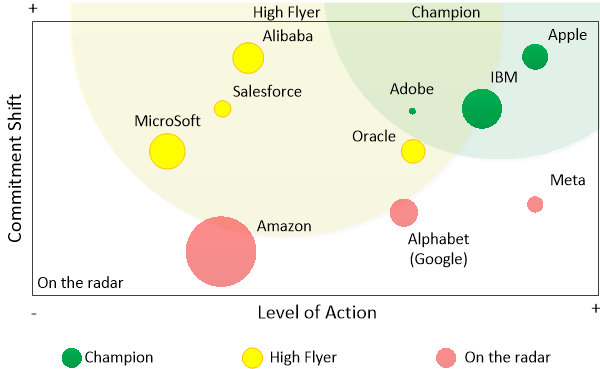}
    \caption{DEI Universe}
    \label{fig:DEIUniverse}
\end{figure*}

Based on our findings, we have also developed a scoring system to position companies. Companies with similar scores were clustered together, indicating shared trajectories within the DEI universe, as shown in Figure \ref{fig:DEIUniverse}. The bubble size represents the most recently reported number of employees; for example, Amazon has the larger bubble. The companies are categorized into three key constellations: Champions (green), High Flyers (yellow), and On the Radar (red).

\section{Discussion}
The spread of DEI in the leading software companies can be summarized as an indicator of the extent and limit of what corporations are willing to do under political and social pressure. To answer our research question \textit{"How have leading software companies changed their DEI strategies in recent years?"}, we summarize our findings in three main points:

\begin{itemize}
    \item Firstly, some companies have significantly scaled back their DEI initiatives, reducing funding or eliminating programs.
    \item Secondly, some companies have renamed their DEI initiatives while some of them have retained their existing activities without major changes or public outcomes. 
    \item Thirdly, some companies have taken clear actions in defending their DEI initiatives and have their stakeholders' support. 
\end{itemize}

Our research, especially our DEI universe map, makes it visible that we are in the middle of crucial changes in DEI initiatives in software engineering. Companies that had previously reported their DEI initiatives and actions, especially in their software product and services design area such as Meta and Google ~\cite{MoisesJaccheri2024, 10.1007/978-3-031-35699-5_25} are, according to our research, decreasing their commitment to DEI. 

DEI is a broad term, and several notions are combined under its umbrella. Still, words matter in this topic. For example, in our research, we identified that some of the companies were renaming their DEI initiatives. In our research, we found that even the whole term "DEI" was somehow challenging. For example, Meta has announced that \textit{"DEI has also become charged"} ~\cite{AllenFischer2025}. However, one could say that in renaming, there may be a risk that replacing the DEI logic with a \textit{"fairness"} or \textit{"justice"} logic may fail to address the underlying structural inequalities. DEI has to evolve, of course, but that does not necessarily mean that we have to remove the term. 

Although backlash against DEI has been identified previously ~\cite{kidder2004backlash, bezrukova2012reviewing}, there were signals that the speed of changes ~\cite{thompson2024coping, sitzmann2024don} was not the only thing to differentiate this wave of DEI backlash from others. The narratives companies used to justify their decisions were mainly political, ideological, and, in some cases, business-driven. For example, Microsoft layoffs were made due to \textit{"changing business needs"}~\cite{Stewart2024}. Google referred to political climate and court decisions \textit{"teams are also evaluating changes to our programs required to comply with recent court decisions and U.S. Executive Orders on this topic."}~\cite{AltchekLangley2025}. Meta also referred to policy change ~\cite{AllenFischer2025}. This aligns with the recent findings that 1 in 8 organizations plan to reduce their DEI commitments in 2025 due to political reasons ~\cite{Resume2025}. Our study shows that this number may become even higher. 

The court decisions are not the only way to interfere with companies' strategies. Stakeholders are one way to affect the DEI strategies, positively or negatively. For example, in Apple's case, the power of stakeholders was shown as DEI negative requests were defended as stakeholders unified and voted against it ~\cite{Apple2025Proxy}. It is also interesting to follow more about what kind of actors create these stakeholder requests, as in actions against the DEI initiative, they have been the same actors in many companies, such as Heritage Foundation and their Project 2025 ~\cite{MacLellan2025AntiDEI,Heritage2025CorporateAccountability}. One important question is whether companies that are not under US court decisions or their stakeholder do not make any anti-DEI requests. Would these companies make some changes to their DEI initiatives in a kind of 'copycat' way, as they feel that they need to react in a similar way?

In this ongoing research, we proposed a DEI universe map to identify the level of actions and community shifts in software engineering companies' DEI initiatives. Our map helps to visualize the changes our industry is going on right now. Our plan is to continue our studies and use the map when investigating more deeply the impact software engineering DEI research had and whether it could create new value, impact, and directions when understanding more about the shift in DEI strategies. The map and our research can give only the idea of what is happening on the surface. In the future, it would be crucial to dive deeper into the companies' actions and content - have only the names of their actions changed or also the actions and in which direction?

Our map also gives signals that not all companies are abandoning DEI initiatives. We hope that this gives strength and motivation for other software engineering companies to make decisions inspired by, for example, 'defending' category companies and follow how they continue developing, creating, and contributing to DEI initiatives in our field. 

\subsection{Threats to validity}

Adams et al. ~\cite{Adams2017ShadesOG} mention several threats to the validity of grey literature reviews, some of which may apply to our analysis as well.

\textbf{Data Source Reliability:} Grey literature often lacks the rigorous peer-review process of academic publications, which can raise concerns about the reliability and credibility of the information. 

\textbf{Selection Bias:} The inclusion of grey literature can introduce selection bias, as the criteria for selecting grey literature may not be as standardized or transparent as those for academic literature.

\textbf{Quality Appraisal:} Evaluating the quality of grey literature is challenging due to its diverse nature and the absence of standardized quality assessment criteria. This can lead to inconsistencies in the quality of included studies.

\textbf{Heterogeneity:} Grey literature is highly heterogeneous, encompassing various types of documents such as reports, theses, and conference papers. This diversity can complicate data synthesis and analysis. Access and Availability: Grey literature is often not as easily accessible as academic literature, which can limit the comprehensiveness of the review and introduce potential gaps in the data.

\textbf{Transparency and Replicability:} The processes for identifying and including grey literature are often less transparent and replicable compared to those for academic literature. This can affect the credibility and reproducibility of the review findings. 

\textbf{Publication Bias:} Including grey literature can help address publication bias, but it also introduces the risk of incorporating low-quality or biased sources that may not have undergone rigorous scrutiny.

In this context, for our study, the most relevant threats are: 
\begin{itemize}
    \item Data Source Reliability: The study relies on grey literature, such as news articles, company reports, and internal memos, which may not always be as reliable or comprehensive as peer-reviewed academic literature.
    
    \item Selection Bias: The selection of the top 10 software companies from the Forbes 2024 Global 2000 list may not represent the entire software industry, potentially leading to biased results.
    
    \item Temporal Limitations: The data collection was conducted in February 2025, focusing on information produced during 2025. This narrow timeframe may not capture longer-term trends or changes in DEI strategies.
    
    \item Political and Social Context: The study is heavily influenced by the political climate in the US, particularly under President Trump's second term. This context may not be applicable to companies operating in different political environments.
    
    \item Interpretation of DEI Initiatives: The categorization of companies into Champions, High Flyers, and On the Radar is based on the authors' interpretation of the data, which may be subjective.
    
    \item Generalizability: The findings may not be generalizable to other industries or regions, as the study focuses specifically on leading software companies.
\end{itemize}

\section{Conclusions}
Diversity, Equity, and Inclusion initiatives have played a significant role in shaping company strategies, especially in software engineering. However, recent shifts especially in the political landscape have revealed that there are changes going on regarding the commitment to the DEI initiatives. To investigate this issue, we conducted a grey literature review for the Top 10 software companies. Our study maps the current DEI landscape in software engineering companies, categorizing them based on their level of action and strategic shifts. Our findings highlight the multifaceted nature of change, where some interests can be advanced while others are left behind. Future research should focus on revisiting the commitment of the main tech companies, in light of further political changes and their implications on funding and industry trends. The DEI universe map presented in this paper reveals that we are in the process of crucial changes regarding DEI initiatives in tech. Therefore, it is vital to keep track of these transformations and see where they lead, both regarding the workforce and software produced. 



\bibliographystyle{ACM-Reference-Format}
\bibliography{software}

\appendix

\clearpage
\section{Appendix 1: Data 2 Grey Literature Review sources}

\renewcommand{\arraystretch}{1.3}
\begin{tabular}{|p{2cm}|p{2cm}|p{10cm}|p{2cm}|}
    \hline
    \textbf{ID} & \textbf{Company} & \textbf{Source Title} & \textbf{URL} \\
    \hline
    22 & Apple & Notice of 2025 Annual Meeting of Shareholders and Proxy Statement & \href{https://www.sec.gov/ix?doc=/Archives/edgar/data/320193/000130817925000008/aapl4359751-def14a.htm}{Link} \\
    19 & IBM & International Business Machines Corporation Shareholder Proposal & \href{https://www.sec.gov/files/corpfin/no-action/14a-8/heritageibm122024-14a8inc.pdf}{Link} \\
    31 & IBM & Complaint and Jury Demand: Randall E. Dill v. IBM & \href{https://media.aflegal.org/wp-content/uploads/2024/08/20173246/ECF001_Dill-v.-IBM_Complaint.pdf}{Link} \\
    16 & IBM & Heritage Foundation on IBM's DEI Requirements & \href{https://fortune.com/2025/02/10/dei-shareholder-proposals-2025-diversity-inclusion-corporate-battle-apple-coca-cola-ibm-berkshire-hathaway/}{Link} \\
    33 & Adobe & The Adobe Foundation and USC Annenberg Inclusion Initiative release new Inclusion List & \href{https://blog.adobe.com/en/publish/2025/01/22/the-adobe-foundation-and-usc-annenberg-inclusion-initiative-release-new-inclusion}{Link} \\
    32 & Adobe & Southern California Wildfires: Supporting the Community & \href{https://blog.adobe.com/en/publish/2025/01/15/southern-california-wildfires-supporting-the-community}{Link} \\
    Footnote 5 & Adobe & Adobe Film \& TV Fund with Group Effort Initiative & \href{https://blog.adobe.com/en/publish/2025/01/22/adobe-film-tv-fund}{Link} \\
    Footnote 6 & Adobe & Adobe Annual DEI Report & \href{https://www.adobe.com/diversity/report.html}{Link} \\
    34 & Microsoft & 2024 Global Diversity \& Inclusion Report & \href{https://www.microsoft.com/en-us/diversity/inside-microsoft/annual-report}{Link} \\
    46 & Microsoft & Microsoft Layoffs of DEI Team & \href{https://www.bloomberg.com/news/articles/2025-01-10/microsoft-dei-team-layoffs}{Link} \\
    3 & Alibaba & 2024 Alibaba Group Environmental, Social and Governance (ESG) Report & \href{https://www.alibabagroup.com/en-US/esg}{Link} \\
    Footnote 7 & Alibaba & Alibaba Global Initiatives - DEI Commitment & \href{https://agi.alibaba.com/home}{Link} \\
    Footnote 8 & Oracle & Oracle DEI Commitment Statement & \href{https://www.oracle.com/corporate/careers/diversity-inclusion/}{Link} \\
    Footnote 9 & Oracle & Oracle Removes DEI Terminology from Website & \href{https://docs.oracle.com/en/industries/financial-services/banking-trade-finance/14.7.5.0.0/prcug/shared-global-topic-diversity-and-inclusion.html}{Link} \\
    28 & Salesforce & Salesforce CEO Interview at World Economic Forum 2025 & \href{https://www.axios.com/2025/01/22/salesforce-chief-ai-agents-davos}{Link} \\
    38 & Salesforce & Salesforce DEI Lawsuit on Free Speech & \href{https://fortune.com/2025/02/01/anti-dei-guide-groups-people-attackin-goldman-ibm-pfizer-jpmorgan-salesforce/}{Link} \\
    12 & Salesforce & Salesforce DEI Initiatives and Future Plans & \href{https://www.salesforce.com/company/dei/}{Link} \\
    36 & Salesforce & Salesforce Diversity Hiring Practices Update & \href{https://www.salesforce.com/diversity-hiring-update}{Link} \\
    Footnote 10 & Salesforce & Internal Memo on DEI Strategy Adjustments & \href{https://www.salesforce.com/internal-dei-memo}{Link} \\
    15 & Amazon & Amazon is Halting Some of Its Diversity and Inclusion Programs & \href{https://www.bloomberg.com/news/articles/2025-01-10/amazon-is-halting-some-of-its-diversity-and-inclusion-programs}{Link} \\
    39 & Amazon & Amazon Memo on Changes to DEI Programs & \href{https://web.archive.org/web/20241126111614/https://www.aboutamazon.com/about-us/our-positions}{Link} \\
    53 & Amazon & Amazon's Revised Workplace Inclusion Policies & \href{https://www.amazon.com/workplace-inclusion-2025}{Link} \\
    Footnote 11 & Amazon & Amazon Leadership Discusses DEI Future & \href{https://www.amazon.com/diversity-inclusion-strategy}{Link} \\
    6 & Alphabet & Google Ends Diversity Hiring Goals - Business Insider & \href{https://www.businessinsider.com/google-ends-diversity-hiring-goals-reviews-dei-programs-2025-2}{Link} \\
    Footnote 12 & Alphabet & Google Removes Heritage and Pride Events from Calendar & \href{https://support.google.com/calendar/thread/322570334/why-is-pride-month-removed-from-the-google-calendar?hl=en}{Link} \\
    55 & Meta & Meta Kills DEI Programs & \href{https://www.axios.com/2025/01/10/meta-dei-programs-employees-trump}{Link} \\
    4 & Meta & Meta CEO Mark Zuckerberg Video on Policy Changes & \href{https://www.axios.com/2025/01/10/meta-dei-memo-employees-programs}{Link} \\
    5 & Meta & Meta Ends Representation Goals for Women and Minorities & \href{https://www.axios.com/2025/01/10/meta-dei-memo-employees-programs}{Link} \\
    \hline
\end{tabular}

\end{document}